\documentstyle[preprint,prb,aps]{revtex}
\draft
\begin{document}
\title{COMPARISON OF THE ELECTRONIC STRUCTURES AND ENERGETICS OF FERROELECTRIC
LiNbO$_{3}$ and LiTaO$_{3}$}
\author{IRIS INBAR}
\author{R.E. COHEN}
\address{Carnegie Institution of Washington, Geophysical Laboratory, \\ 5251
Broad Branch Rd, N.W. Washington, D.C. 20015}
\date{\today}
\maketitle
\begin{abstract}

  Extensive LAPW frozen phonon calculations were performed in order to
 understand the origin of ferroelectricity in LiTaO$_{3}$ and LiNbO$_{3}$.
 Displacement of the Li atoms alone results in an anharmonic single well,
 whereas displacements of oxygen and lithium together result in deep
 double wells, much deeper than the transition temperatures, T$_C$.
 This is contrary to current theories which model the underlying potential
 as a triple well potential for the lithium atoms.
 Our results support an order-disorder model for the oxygen atoms as
 the driving mechanism for the ferroelectric instability.
 Oxygen displacements alone against the transition metal atoms result in
 shallower double wells as a result of oxygen-lithium overlap so that
 the lithium and oxygen displacements are strongly coupled .  We find
 large hybridization between the oxygens and the
 transition metal atoms. Thus ferroelectricity in the Li(Nb,Ta)O$_{3}$ system
 is similar in origin to ferroelectricity in the perovskites.
  We also find that the electronic structures of LiTaO$_{3}$ and LiNbO$_{3}$
 are very similar and hardly change during the phase transition.

\end{abstract}


  \section{Introduction}

  The origin of ferroelectricity in the two well known ferroelectric systems,
 LiNbO$_3$ and LiTaO$_3$, has been subject to intense study  since the
 discovery of LiNbO$_{3}$ in 1949 \cite{mr}.  They have
 many applications in optical, electro-optical and
 piezoelectric devices, but the fundamental physics that leads to their
 ferroelectric behavior have not been studied.  Their transition temperatures,
 which are among the highest known ferroelectric transition temperatures,
 are quite different, 1480 K for LiNbO$_{3}$ and 950 K for LiTaO$_{3}$.
 The electronic origin of their different $T_{C}$ is a mystery since
 Nb$^{5+}$ and Ta$^{5+}$ behave very similarly, and structurely these
 materials are almost identical.
 The origin of their ferroelectric instability as well as their
 different transition temperatures is investigated.

 Both materials undergo only one structural phase transition.
 The paraelectric structure has a 10 atom unit cell and the average
 structure belongs to the R$\bar{3}$c space group,  The atomic arrangement
 consists of oxygen octahedra sharing faces along the polar trigonal
 axis.  The transition metal atoms occupy the centers of oxygen octahedra,
 and the average Li atom position lies on the face between two adjacent oxygen
 octahedra (Fig. \ref{structure}a).
 The ferroelectric structure is rhombohedral, and belongs to the space
 group R3c.  The transition metal atom is displaced from the center of
 the oxygen octahedra along the trigonal axis. The next oxygen octahedron
 along this axis is empty and the adjacent octahedron has a Li atom
 ferroelectrically displaced from the oxygen face in the spontaneous
 polarization P$_s$ direction (Fig. \ref{structure}b).
 Glass in 1968 and later Johnston {\it et al.} have determined,
 using dielectric and thermal measurements, that the phase transformation
 in these systems is continuous \cite{gl68,johnston}.

 Whether the transition is displacive or order-disorder has been much
 discussed and confusion abounds.
 A displacive phase transition is one where the local potential in the
 mean field of the rest of the crystal has a single
 minima, and is characterized by a temperature dependent optic mode
 approaching zero as the temperature reaches T$_{C}$.
 Temperature dependence measurements of Raman, Rayleigh scattering
 \cite{johnston} and infrared reflectivity \cite{ser,bark1,bark2}
 show soft mode behavior for one polar A$_{1}$ (TO) optic mode in the
 ferroelectric phase.  This soft mode crosses many E modes
 (whose eigenvectors give ionic displacements perpendicular to the polar axis)
, and thus
  this specific mode is  difficult to trace in detail.
 Tomeno and Matsumura \cite{tomeno} measured the dielectric constants
 of LiTaO$_{3}$ and found a large Curie constant, and interpreted their
 results as indicative of a displacive transition.

 Evidence for the transition having an order-disorder character came from
 Penna {\it et al.} \cite{pen1,pen2,pen3,pen4}, who observed no mode softening
 for LiTaO$_{3}$ for the A$_1$ (TO) mode, followed by Chowdhury {\it et al.}
 \cite{chow} who performed neutron scattering experiments on LiNbO$_3$, and
 also failed to observe any softening of the A$_1$ mode.  Okamoto {\it et al.}
 \cite{oka} used Raman scattering to study LiNbO$_3$ between
 room temperature and 1225K and saw two of the three A$_1$
 modes (which are, at room temperature, TO, LO and TO at 250, 270 and 274
 cm$^{-1}$, respectively), and observed anomalous behavior for one of
 them (at 274 cm$^{-1}$) as the temperature reached T$_c$.
 They noted that the decrease in the peak frequency was mostly due to the
 rapid increase in damping as the linewidth had a divergent form in
 temperature, whereas the
 quasiharmonic frequency remained almost temperature-independent.
 Their conclusion was therefore that LiNbO$_3$
 does not exhibit a typical displacive transition, but rather resembles
 an order-disorder system.  Zhang {\it et al.}
 \cite {zhang} reported these measurements on LiTaO$_3$, and found the same
 kind of behavior.   In an order-disorder phase transition the local
 potential is characterized by a double (or more) well, with the thermal energy
 $kT_{C}$ much smaller then the well depth and no soft phonon mode
 exist since phonons now oscillate within each well and the wells remain
 essentially unchanged throughout the phase transition. These transitions
 are characterized by a diffusive soft mode which is not a phonon but
 represent large amplitude thermal hopping between the wells.
 At T above $T_{C}$ the crystal is non-polar in a thermally averaged
 sense.

  Jayaraman {\it et al.} \cite{jaya} argued for an order-disorder type
 transition because they saw little pressure dependence of the Raman mode;
 they also emphasize the difference from ferroelectric perovskites, which
 show a strong
 pressure dependence.  Raptis \cite {cons}  measured and analyzed Raman
 modes of LiTaO$_3$ between room temperature and 1200K and observed
 softening of this A$_1$ mode (along with others) to a certain degree.
  However, the decrease was characterized with an order-disorder model.
 Catchen {\it et al.} \cite{catch} used perturbed-angular-correlation
 (PAC) spectroscopy to measure nuclei-electric-quadrupole interactions at
 the Li sites over a temperature range of 295-1100K, and
 Cheng {\it et al.} f\cite {cheng} studied inelastic neutron scattering
 from room temperature up to 800$^{0}C$ (1100 K) in LiTaO$_3$; both failed
 to observe mode softening, therefore not supporting the
 displacive  picture for the phase transition.
 Tezuka {\it et al.} \cite{tezuka} used hyper-Raman and Raman spectra
 of LiTaO$_{3}$ between 14 and 1200 K.  No evidence was found for the
 softening of the $A_{1}$  mode; however, a strong Debye-type relaxational
 mode was found in the two phases, suggesting an order-disorder type
 transition.  They interpreted the anomalous line shape of an $A_{1}$ mode
 in terms of coupling with relaxational modes.

 Most ferroelectric systems are thought of as exhibiting displacive
 behavior far from the transition temperature region and order-disorder
 characteristics near T$_C$.   In the Li(Nb,Ta)O$_3$ systems this conclusion
 is supported by a number of studies \cite{samu,prieto,tome}.

 In other ferroelectric oxides, like the perovskites KTaO$_{3}$ and KNbO$_{3}$,
 the mechanism behind the phase transition has also been debated. Evidence
 for the transition being of the displacive type are presented by Nunes
 {\it et al.} \cite{nunes} and Samara \cite{sama}, while Comes {\it et al.}
 \cite{comes} and PAC experiments by Dougherty {\it et al.} \cite{keith}
 point to it being of the order-disorder type.  Sokoloff {\it et al.}
 \cite{sokoloff} have studied Raman scattering of KNbO$_{3}$ and BaTiO$_{3}$
 and discovered central peaks which have line shapes and thermal dependence
 characteristics of Debye relaxation modes as well as symmetry properties
 consistent with the eight-site model.  A theoretical
 study by Edwardson of KNbO$_{3}$ \cite{ed}, using interacting polarizable
 ions in static and dynamic simulations found a mixture of order-disorder
 and displacive type behavior.  Postnikov {\it et al.} \cite{post1,post2}
 carried out an LMTO study examining the total energy of KNbO$_{3}$ in the
 tetragonal and rhombohedral phases. These calculations found
 that displacing the Nb atom along the $<$100$>$ direction from its tetragonal
 cell position
 corresponds to a saddle point on the total energy surface. This becomes
 a minima when the lattice strain is included.  They have also carried out
 total energy calculations for KTaO$_{3}$ which found no ferroelectric
 instability for the calculated volume.  They were able to induce a phase
 transition by applying negative pressure (expanding the lattice).
 An XAFS study of KTN by Hanske-Petitpierre {\it et al.} \cite{hanske}
 found that the ferroelectric transition is not displacive and involves
 orientational order-disorder transition of the Nb atom.
  A recent first principles investigation of eight perovskites
 \cite{king,kisas} suggests that in materials like KNbO$_{3}$ and
 BaTiO$_{3}$, which are rhombohedral at T=0, the sequence of successive
 transitions is explained via the eight site model, where the order
 parameter in the paraelectric phase fluctuates between the eight minima
 in the [111] directions.  These sites are minima at the cubic phase,
 before the development of the strain.

 The theories developed thus far for the LiTaO$_{3}$ and LiNbO$_{3}$ systems
 usually are based on the Lines model \cite{lines}.
 M. E. Lines applied his effective-field theory to
 LiTaO$_3$, and parametrized it as a displacive ferroelectric due
 to the data available at the time by Johnston and Kaminow \cite{johnston},
 and assumed a triple well potential for the Li atoms.   Abrahams {\it et al.}
 \cite{abrams1}
 performed neutron scattering of LiTaO$_3$ between room temperature and
 940~K, and discovered that, above T$_C$, the lithium atom positions in
 LiTaO$_3$ become disordered and hop among the centrosymmetric position
 and sites at $\pm$ 0.37~\AA~  along the optic axis. Similar measurements
 for LiNbO$_3$ show the same kind of behavior \cite{boby}.  The neutron
 scattering data is the cornerstone behind all theories modeling
 this ferroelectric transition as an order-disorder mechanism with the
 Li ions hopping among the centrosymmetric sites and the adjacent
 octahedral sites.  This approach was adopted by Birnie
 \cite{birnie1,birnie2,birnie3}, who modeled the Li hopping as a Frenkel
 defect, and later by Bakker {\it et al.} \cite{bakker}, who used this data
 in addition  to the triple well Lines model as a basis for
 a quantum-mechanical description of the phase transformation in LiTaO$_3$.
 Bakker {\it et al.} predicted and observed \cite{bakker1} a 32 cm$^{-1}$
 excitation which they ascribed to Li motions between the central
 and lowest wells. This excitation frequency has not been observed in
 other studies, however \cite{nelson}.

  \section{Method}

 The Kohn-Sham equations \cite{ksh,ksh1} are solved self-consistently using the
 full potential Linearized Augmented Plane Wave (LAPW) method \cite{hk1},
 where the electronic many-body exchange-correlation interactions are
 described by the local density approximation (LDA) using Hedin-Lundqvist
 parametrization \cite{hl}.
 There are no shape approximations for the charge density or the potential.
 This method has proved predictive in many previous studies.  Examples
 include the prediction of a high pressure phase transformation in
 silica \cite{kingma}, studies of iron at high pressures \cite{stixr,stix1}
 studies of Al$_{2}$O$_{3}$ \cite{freddi}, MgO and CaO \cite{mehl},
 MgSiO$_{3}$ \cite{ronmso} and high temperature superconductors \cite{compu}.
 This method was previously applied successfully to perovskite ferroelectrics
 like BaTiO$_{3}$, PbTiO$_{3}$ \cite{ronnie} and KNbO$_{3}$ \cite{sinboy}.

 We use the LAPW+LO method \cite{singh} which uses a mixed basis consisting of
 the LAPW basis plus extra localized orbitals inside the muffin tin spheres.
 The extra local orbitals remove a Li s ghost state and
 relax the valence states.  It also allows the use of a single energy window.
 Local orbitals included s for Li, {\it s} and {\it p} for
 O and {\it s}, {\it p} and {\it d} for the Nb and Ta atoms.  Other details
 of the calculations include a muffin tin size of 1.6 Bohr for the Li and 2.0
 Bohr for the Nb and Ta atoms. The oxygen's muffin tin radius was 1.6 Bohr
 for oxygen-Nb separation up to 1.882\AA~ (3.556 Bohr). The corresponding
 distance in LiTaO$_3$ is O-Ta up to 1.8845\AA~ (3.562 Bohr).   At this point
 the oxygen muffin tin radius was decreased to 1.552 Bohr.  For smaller
 separations the muffin tin radius was 1.506 Bohr.  In order to be able to
 compare the energies calculated using different muffin tin radii, we have
 repeated calculations with the three sets of muffin tin radii to find the
 energy shift due to this change in the muffin tin radii, and have assumed
 that this small shift ($\sim$ 7 mRy) is constant for small displacements of
 atoms.

  A 4x4x4 special k-point mesh was used which generates a total of 10
 k-points in the irreducible zone.
 To test energy convergence, the energies at the symmetric and experimental
 structures of both LiTaO$_3$ and LiNbO$_{3}$ were also calculated with
 a 6x6x6 mesh  which generates 28 k-points in the irreducible
 zone, and these energies are shown in
 Table \ref{ene}.  The change in energy difference for the two k-point sets
 between the experimental and the symmetric configurations is 0.069 mRy for
 LiNbO$_{3}$ and 0.3 mRy for LiTaO$_{3}$, demonstrating convergence.

 The RK$_{max}$ parameter was set to 7.0 which gives approximately
 1150 basis functions for the LiTaO$_3$ calculations and
 1050 functions for LiNbO$_3$.
 The core states were calculated fully relativistically and
 the valence states semi-relativistically.
 For each Ta atom, the states up to 4f  were included in the core,
 and as a result 0.588 electrons extended beyond the muffin
 tin sphere.
 For each Nb atom states up to 4s were included in the core and only
 0.07 core electrons extended beyond the sphere; core electrons that
 spill out of the muffin tins see an extrapolated spherical core potential.
 Also, in LiNbO$_{3}$ the Li atom s states were included as bands, whereas in
 LiTaO$_{3}$ they were treated as core states.

 \section{Results and Discussion}
 \subsection{Energetics}

 We have calculated the potential energy surfaces along the experimental
 soft mode coordinate.
 To test the sensitivity to the different lattice parameters, the total energy
 of LiNbO$_3$ was calculated in the ferroelectric configuration using both
 the LiNbO$_3$ lattice parameters (a$_H$=5.14829\AA~  and c$_H$=13.8631\AA~)
 \cite{abrams2}, and the LiTaO$_{3}$ lattice parameters
 (a$_H$=5.15428 \AA~ and c$_H$=13.78351 \AA~) \cite{abrams1}, a difference
 of 0.75\% in the c/a ratio.
 The effect of this strain on the total energies was almost negligible;
 slightly less then 1 mRy or 5.5\% of the well depth.  This is in contrast
 to the case of the perovskites; e.g. PbTiO$_3$, where a strong dependence
 of the total energy on the tetragonal strain was observed, and the energy
 decreases markedly, about 35\% of the well depth for the experimental 6\%
 c/a strain \cite{ronnie}.
 Also the total energy of LiNbO$_{3}$ using both the experimental LiNbO$_3$
 atomic positions (Li at (0.2829,0.2829,0.2829), Nb at (0,0,0) and Oxygen at
 (0.1139,0.3601,-0.2799)) \cite{abrams2} and the LiTaO$_{3}$ atomic
 positions (Li at (0.279,0.279,0.279), Ta at (0,0,0) and Oxygen at
(0.1188,0.3622,-0.2749) \cite{abrams1}) were calculated.  The resulting wells
 are less then 1 mRy different, the LiNbO$_{3}$ atomic positions yielding
 the deeper well. For the purpose of comparison, except for a few more points
 which yielded the same results (energy differences of less then 1 mRy), all
 points were calculated using the experimental lattice parameters and positions
 of LiTaO$_{3}$ to facilitate comparison of the effects of chemistry on
 ferroelectric behavior and electronic structure.  Table \ref{ene} summarizes
 the results for LiTaO$_3$ and
 LiNbO$_3$. The first column refers to the displaced atoms, and to the amount
 of
 displacement as a fraction of the paraelectric to experimental ferroelectric
 normal mode amplitude.  Note that one distortion of LiTaO$_{3}$ published
 earlier \cite{roli} was not along the soft mode coordinate; the present
 results correct this error.

  Figure \ref{energy}a shows the potential energy surfaces of LiTaO$_3$
 with respect to displacements of Li only (upper curves), oxygen only (middle,
 shallow double wells) and Li+O (lower curves).
  Figure \ref{energy}b shows the same picture for LiNbO$_{3}$.
 The lithium displacements along the soft mode coordinate result in a single
 anharmonic well with low curvature.  Displacing only the oxygens against the
 transition metal atoms results in shallow double wells, and the deep double
 wells are the result of the Li+O displacements along the experimental
 ferroelectric coordinate.

 The wells resulting from the oxygen and lithium
 displacements have well depths of 17.3 mRy (2739 K) and 18.3 mRy
 (2858 K) for LiTaO$_3$ and LiNbO$_3$, respectively.
  Both wells are much deeper than the experimental transition
 temperatures, which is consistent with an order-disorder character for the
 phase transition.

 The energy was fit to a fourth order polynomial in normal mode
 amplitude, $ Q = \sqrt {\sum_{i} {m _{i} u_{i}^{2}}} $.
 The Schr\"odinger equation was solved numerically to obtain the eigenstates
 assuming  one-dimensional non-interacting anharmonic
 oscillators along the soft-mode coordinate.
 Figure \ref{levels} shows the energy surface as a function of the normal mode
 amplitude, Q, and the energy levels.
 We can see that the two wells have a different shape due to the factor of
 about two in mass of Nb and Ta, and thus the Nb approximately displaces twice
 as much as the Ta relative to the center of mass, which results in
 a different normal mode amplitude Q.
 The energy difference between the ground and lowest excited state gives
 a frequency for LiTaO$_3$ of 270 cm$^{-1}$ in fairly good agreement with the
 experimental Raman frequency of 201-225 cm$^{-1}$
 considering the one dimensional noninteracting oscillators approximation.
 For LiNbO$_3$, the calculated frequency is  250 cm$^{-1}$,
 in excellent agreement with experimental data of about 250-275 cm$^{-1}$.

 These results indicate that these structural phase transitions are {\it not}
 dominated energetically by the displacements of the lithium alone.
 The potential energy surfaces show that the deep double wells are the
 result of the {\it coupled} motion of lithiums and oxygens. Displacement
 of the lithiums alone hardly changes the energy of the system. This
 is in contrast to current theories which model the displacement of the
 lithiums as the driving mechanism for the ferroelectric instability.

 In order to understand the oxygen-lithium  coupling,  we calculated
 the dynamical matrices for the LAPW and Madelung energies (assuming fully
 charged ions) for both materials.
 The LAPW and Madelung energies were fitted to a 4th order polynomial surfaces
 in the normal mode coordinates of the lithium and oxygen (Q$_{Li}$ and
 Q$_{O}$).
  The second derivatives of these energy surfaces at zero
 displacements are the coefficients of the dynamical matrices.
 Table \ref{coef} shows the coefficients of the fit for the total energy of
 LiNbO$_{3}$ and LiTaO$_{3}$; all the coefficients for LiNbO$_{3}$ and
 LiTaO$_{3}$ are well constrained
 except the coefficient of $Q_{Li}^{2}$, which means the potential surface
 describing the displacement of the lithium only could be either a single
 anharmonic well or a very shallow double well (corresponding to a positive
 or a negative sign).
 Linear and cubic terms (e.g;  Q and $Q^{3}$) are excluded from the fit
 based on symmetry considerations, and terms which are not along the
 coordinates calculated; $Q_{Li}$, $Q_{O}$ and $Q_{Li,O}$, are excluded from
 the fit since they degrade the variances of the quadratic coefficient. These
 include terms like $Q_{Li}^{3}Q_{O}$ and $Q_{O}^{3}Q_{Li}$.
 The dynamical matrix in units of $Ryd^2 \AA^{-2} amu$ representing the LAPW
 energies of LiNbO$_{3}$ and LiTaO$_{3}$ are
 $$ D_{LAPW}(LiNbO_{3})= \left( \begin{array}{cc}
    D_{Li} & D_{Li,O} \\
    D_{Li,O} & D_{O}
 \end{array} \right) =
 \left( \begin{array}{cc}
 -0.001 & -0.012 \\
 -0.012 & -0.015
 \end{array} \right)  $$
 $$ D_{LAPW}(LiTaO_{3})= \left( \begin{array}{cc}
 0.0018 & -0.012 \\
 -0.012 & -0.023
 \end{array} \right).  $$
 The lithium-only contributions ($D_{Li}$) are the smallest (an order
 of magnitude smaller then the rest), the oxygens only contributions are
 larger, with $D_{O}$ in LiTaO$_{3}$ larger than LiNbO$_{3}$.

  Whether the origin of lithium and oxygen coupling is Coulombic
 can be determined by looking at the Madelung contribution to the
 dynamical matrices.
 The Madelung energies were calculated using experimental positions and
 lattice parameters and full ionic charges.  The second derivatives at
 zero displacements, which are the elements in the dynamical matrix, were
 calculated numerically and are shown in Table \ref{coef}.
 In the case of LiNbO$_{3}$ and LiTaO$_{3}$ the Madelung contributions to the
 dynamical matrix are
 $$ D_{MAD}(LiNbO_{3})= \left( \begin{array}{cc}
  0.038 & 0.004 \\
  0.004 & -0.171
 \end{array} \right)  $$
 $$ D_{MAD}(LiTaO_{3})= \left( \begin{array}{cc}
 0.04 & 0.003 \\
 0.003 & -0.162
 \end{array} \right).  $$

  As expected, $D_{O}$ has the largest magnitude,
 followed by $D_{Li}$.  The coupling term between the lithiums and the
 oxygens is in fact zero.  This means that the origin of the lithium-oxygen
 coupling is {\it not} pure Coulombic (Madelung).

 Another possibility is the polarization of the oxygens by the lithium
 displacement, leading to changes in Nb(Ta)-O bonding. The experimental
 ferroelectric configuration generates an
 effective dipole at the lithium sites.  This dipole field can polarize the
 oxygens and drive them off center, yielding a ferroelectric distortion.
 We have plotted the self-consistent charge densities in two configurations;
 in one only the oxygens are displaced and in the other both the oxygens and
 the lithiums are displaced.   In order to see the effects of displacing the
 lithiums we subtracted the two charge densities.  This is shown in
 Fig. \ref{polarization} where the charge density contours are plotted
 on a scale of -0.1 to 0.1 electrons/bohr$^{3}$ and the contour interval
 is 0.002 electrons/bohr$^{3}$.  A large dipole is
 seen at the lithium sites due to the displacement of the lithiums. Little
 polarization of the oxygens is observed; there is no evidence for any large
 dynamical covalency effects.
 We can therefore eliminate the possibility of oxygen-lithium coupling through
 either Madelung or polarization effects.

 Another possible source for the oxygen-lithium coupling is through the
 crystal structure.
 It is important to notice that the oxygens move not only along the c axis,
 but rather have sizable displacements along the a and b axes as well.
 We have tested the importance of
 these displacements by moving only the oxygens along the c component of the
 experimental ferroelectric displacement (the polar axis).
 The resulting energy curve was far shallower then the energy surface
 which resulted from moving the oxygens only along the experimental
 soft mode coordinate. This is shown in Fig. \ref{zonly} where the upper
 curve represents the displacements of the oxygens along the polar axis only
 and the lower, deeper well represents the total energy when displacing the
 oxygens along the experimental ferroelectric distortion (along a,b and c
 axes).  The reason for these big energy differences can be seen from Table
 \ref{bonds} which shows the Ta-O, Nb-O and Li-O bond lengths.  The ionic
 radii of Li is about 0.6 \AA~, that of Ta or Nb is about 0.6 \AA~ and
 the ionic radii of oxygen is about 1.4 \AA~ \cite{ionicrad}, making the sum
 of each pair (Li-O, Nb-O and Ta-O) about 2.0 \AA~.
 When the oxygens are displaced only along the c axis, the oxygen-Nb (Ta)
 separation becomes only 1.83 (1.86) \AA~ which is about 0.17 (0.14) \AA~
 shorter then the sum of the ionic radii (about 8 (7)\%).  Therefore it is
 energetically favorable for the oxygens to displace in the a-b plane as
 they move along the c axis.

  If we now consider the experimental ferroelectric displacement of only
 the oxygens and the experimental ferroelectric {\it coupled} displacement of
 the oxygens and lithiums,  displacing the oxygens only
 results in a Li-O separation which is also shorter than the sum of their
 ionic radii (Table \ref{bonds}).   This explains why the wells
 associated with the oxygen displacements alone are shallower than those
 obtained with the displacement of {\it both} the lithiums and oxygens.
 The origin of the Li-O coupling is therefore the fact that motion of the
 oxygens alone yields a Li-O distance that is larger then the sum of their
 ionic radii, resulting in a deeper well for the coupled motion (in which
 the Li and oxygens move away from each other).

 We can therefore conclude that the driving mechanism behind the phase
 transformation in these systems is the displacement of the oxygens towards
 the transition metal atoms.
  Displacement of the oxygens in the direction of
 the transition metal atoms only (the c axis) would result in too short
 Nb (Ta)-oxygen bonds. The oxygens therefore move also in the
 plane perpendicular to the c axis, towards the lithiums.  This shortens the
 lithium-oxygen bond so that the lithium displacements
 are coupled with the oxygen motions.

 The transition temperature, T$_{C}$, cannot be calculated directly from the
 zone center energetics.  In the usual models for ferroelectric phase
 transitions, T$_{C}$ is related to the relative strength of the local
 (on-site) and coupling terms in the energy \cite{kissas1}.  Since we find
 the zone center energetics to be similar, the difference in T$_{C}$ must be
 due to differences in the energetics at the zone boundary.\\
  In order to understand the origin of the ferroelectric distortion, we next
 examine the electronic structure of these materials.

 \subsection{Electronic Structure}

  One goal of this research is to understand the origin of
 ferroelectricity in LiTaO$_{3}$ and LiNbO$_{3}$ and the difference in
 $T_{C}$ from their electronic structure.
  Fig. \ref{dos2}a compares the electronic density of states for LiTaO$_{3}$
 and LiNbO$_{3}$, both at the ferroelectric configuration.   The energy
 scales are lined up with the top of the valence bands at zero energy.
 It is clear that the total density of states of these two materials is
 very similar.  We can look further at the different contributions to the
 density of states;  Fig. \ref{dos2}b compares the partial density of the Ta 5d
 state and the Nb 4d state both in the  ferroelectric phase.   The top of the
 valence band is composed mostly of oxygen p states.  This figure shows
 a large density of transition metal d states in the valence band which means
 that the oxygen p states in these two materials are hybridized with the d
 states.  The Nb d
 states have a large peak at the bottom of this band which is missing in
 the case of the 5d states of the Ta atom.  The origin of this peak is the
 fact that the lowest valence bands of LiNbO$_{3}$ (the bands at about -4.5 eV
 or -0.35 Ryd) are less dispersive than the lowest valence bands of
 LiTaO$_{3}$. This will be further discussed later.
  The same conclusion is derived from Fig. \ref{dos2}c of the
 partial density of the oxygen p state of the two materials in the valence
 band in the ferroelectric structure.  Here too,
 the densities of states are very similar.  The same peak at the bottom of the
 band is seen here for the LiNbO$_{3}$, which is missing in the valence band
 of LiTaO$_{3}$.
 All three figures that compare the total and partial density of states of the
 two materials in their ferroelectric phase show  large hybridization
 between the transition metal d states and the oxygen p states, which is
 the reason for the oxygen displacements towards the transition metal atoms.

 Next we compare the densities of states in the paraelectric
 and the ferroelectric phases.
 Fig. \ref{dos1}a illustrates the total density of states of LiNbO$_{3}$
 in the paraelectric (solid line) and the ferroelectric (dashed line) phases.
 The bands in the two phases look similar except that the bands at the
 ferroelectric phase are slightly wider then the bands at the paraelectric
 phase.
 Fig. \ref{dos1}b compares the Nb 4d state both in the paraelectric (solid
 line) and the ferroelectric (dashed line) phases and Fig. \ref{dos1}c shows
 the Ta 5d states in the
 two configurations.  The large peak at the lower part
 of this band (the peak at about -0.35 Ryd or -4.5 eV) is shifted in the
 ferroelectric
 case to higher energies.
 Fig. \ref{dos1}d shows the density of oxygen p states of LiNbO$_{3}$ at the
 two phases.  The peak at the bottom of the band is shifted in the
 ferroelectric phase from the paraelectric one.
 These figures indicate that the electronic structure at the paraelectric
 and the ferroelectric phases are quite similar.
  Fig.\ref{dos1}e  compares the lithium 2s character in the paraelectric
 and ferroelectric phases.  It is evident that the lithium is almost
 completely ionized and that its electronic distribution does not change
 during the phase transition.

 Fig. \ref{bands} shows the band structure of LiNbO$_{3}$ in the
 ferroelectric state.
 The band gap is indirect, the top of the valence band is between $\Gamma$
 and Z, and the bottom of the conduction band is at the $\Gamma$ point.
 The Brillouin zone for the rhombohedral lattice \cite{koster} is illustrated
 in the inset in the figure.  The energy between the Z and the A point was
 calculated along
 a straight line between the two points not along the Brillouin zone face
 for the purpose of comparison with Ching {\it et al.}'s results
 \cite{ching}.

 The band gap is 3.1 eV which is about 15 \% lower than the value obtained
 from optical measurements of the near stoichiometric sample \cite{gap1,gap2}
 of 3.78 eV.
 The lithium 2s states are separated by 13.6 eV from the oxygen 2s states.
 These bands would not appear in the LiTaO$_{3}$ band structure since the
 lithium 2s states were treated as core states; these bands are very flat.
 The oxygen 2s are separated by  10 eV from the valence bands.
 The lowest conduction bands are the Nb 4d states for LiNbO$_{3}$ or Ta 5d
 states for LiTaO$_{3}$.
 We have also compared this band structure with Ching {\it et al.}
 \cite{ching}  who used the OLCAO method and got a band gap
 of 3.56 eV and the bands compare well with our results.

 Fig. \ref{bands1}a and b show the band structure for LiTaO$_{3}$ and
 LiNbO$_{3}$, respectively.  Each figure shows the ferroelectric (solid line)
 and the paraelectric (dashed line) phases.
 The changes observed between the ferroelectric and paraelectric bands are
 the band gap which is larger in the ferroelectric phase by about 15 \%
 and the band width, especially the conduction band, which is larger in
 the paraelectric phase.

  The band structures of LiTaO$_{3}$ (dashed line) and LiNbO$_{3}$ (solid
 line) both in the ferroelectric phase, are shown in Fig. \ref{bancom}.
 The only difference between these two band structures is the larger
 band gap, by about 1 eV (30 \%), in LiTaO$_{3}$. The conduction bands
 are shifted by 1 eV from each other, but otherwise, their structure is
 almost the same.
 The valence bands are almost identical, which is consistent with the
 results of the total energy calculations where the two well depths were
 found to be very close to each other (within 1.2 mRy of each other) and
 the fact that the number of valence electrons in the muffin tins in both
 materials was similar.
 The only difference which was found between the two electronic structures
 was in the size of the band gap and the less dispersive nature of one band.
 The difference in the band gap will have an effect on quantities which
 include summing over unoccupied states as well as the occupied ones, like
  the polarizability.  This difference will lead to different phonon
 dispersion in the two materials, and thus to different T$_{C}$'s. Zone
 boundary or linear response calculations are necessary to further explore
 this issue.

  \subsection{Comparison to the Perovskites}

 Previously we studied the difference between  the self-consistent charge
 densities and charge densities computed using overlapping ions with the PIB
 model for both LiTaO$_{3}$ and LiNbO$_{3}$ in the ferroelectric phase
 \cite{us}.  In the PIB model which is a non-empirical ionic
 model \cite{pib}, the charge densities are calculated via a Gordon-Kim
 type model \cite{gk}, where the ions are allowed to breath corresponding
 to changes in the crystal potentials.  The comparison indicated large
 hybridizations between the Ta atoms and the oxygens and between the Nb atoms
 and its oxygen neighbors and the Li atoms were fully ionized in the
 self-consistent charge density.

 These results are consistent with the energetics and electronic structures
 results, all pointing to the same conclusion that
 the driving mechanism behind the ferroelectric
 instability in the LiNb(Ta)O$_{3}$ systems is the hybridization between
 the d states on the transition metal atoms and the 2p states on the
 oxygens. The lithiums are but passive players in the ferroelectric
 instability.  This is very similar to the ferroelectric mechanism in
 the perovskite ferroelectrics, where the oxygen-transition metal atom
 hybridization, in addition to the Coulombic long range interaction
 which tend to drive the system off center, overcome the short range
 repulsions which tend to leave the system in it's high-symmetry configuration.

 An interesting comparison can be made with the K(Nb,Ta)O$_3$ system;
 one major difference  is the fact that the perovskite
 KTaO$_3$ is an incipient ferroelectric where LiTaO$_{3}$ has a high
 transition temperature.  In this sense a qualitative
 comparison can be made between the two sets of systems, as in both
 systems the transition temperature is higher in the niobate systems,
 being zero for KTaO$_{3}$.  This would mean a shallower well for the
 tantalates, where in the case of KTaO$_{3}$ the well is apparently
 lower than the thermal vibrations, as shown by this study for the
 Li(Nb,Ta)O$_{3}$ systems and by Postnikov for the K(Nb,Ta)O$_{3}$ systems
 \cite{post1}.

 The electronic structures of the two sets of systems (LiNb(Ta)O$_{3}$ and
 KNb(Ta)O$_{3}$) show a large hybridization between the transition metal
 atoms and the oxygens, and the amount of hybridization between the
 transition metals and the oxygens in the two sets of systems is similar.
 This can be seen from  Fig. \ref{dos2}b  which shows  both the
 partial density of Nb 4d states and that of Ta 5d states in the valence
 bands in the ferroelectric configuration in the LiNb(Ta)O$_{3}$ systems.
 We can compare these results to the same densities of states calculated for
 KNbO$_{3}$ and KTaO$_{3}$ by LMTO \cite{post1} reproduced in
 Fig. \ref{pos}.
 These figures show the Nb (Ta) density of states with the Nb (Ta) atom
 undisplaced and  displaced by 0.073a (a being the lattice constant) along
 the $<$111$>$ direction, which is an exaggerated displacement used to
 enhance the differences between the two phases.
 We can see that the same trends exists in this picture
 as in Fig. \ref{dos2}b for the Nb (Ta) in LiNbO$_{3}$ (LiTaO$_{3}$).   The
 ferroelectric peaks at the bottom of this band are slightly shifted towards
 higher energies.
  In both cases there is large hybridization between the transition metal
 atoms and the
 oxygens, but this hybridization does not change much during the transition.
 This hybridization is essential for the onset of the ferroelectric
 instability, however the amount of hybridization in this system
 doesn't change much through the phase transition like in other ferroelectrics,
 e.g BaTiO$_{3}$ or PbTiO$_{3}$ \cite{ronnie}.

 We find that LiNbO$_{3}$ and LiTaO$_{3}$ are almost identical in their
 electronic behavior. The amount of d character in the valence bands, which
 is a measure of the hybridization between the transition metal ions and
 their oxygen neighbors is very similar.  This is in contrast to the
 conclusion that Ta is less ionic than Nb, reached by Postnikov
 {\it et al.} for K(Ta,Nb)O$_{3}$. This conclusion was based on the smaller
 transition metal atom density of states peak at the bottom of the valence
 band, observed in Figures \ref{pos}a and b.  It is seen from Fig. \ref{bancom}
 that the origin of this peak is the less dispersive nature of one
 LiTaO$_{3}$ band versus LiNbO$_{3}$.

 The similarity between LiNb(Ta)O$_{3}$ and KNb(Ta)$_{3}$ is the fact that
 the driving mechanism for the phase transition in the two systems
 is oxygen-B atom hybridization.  The difference between the two systems
 lies in the different structure which yields a different oxygen-A atom
 interaction.

 In both LiNb(Ta)O$_{3}$ and KNb(Ta)O$_{3}$  the hybridization
 between the B atoms (the transition metals) and the oxygens causes the
 oxygens and the B atoms to displace towards each other.
 In the LiNb(Ta)O$_{3}$ system the oxygen-B atom separation is
 larger than the sum of their ionic radii and the oxygens markedly displace
 in the a-b plane as they move along the polar axis.  This however, makes the
 oxygen-A atom separation larger then the sum of their ionic radii
 resulting in the coupled oxygen-A atom motion.  This is in contrast to the
 perovskites where the A site is large enough to allow the oxygens to move
 towards the A atoms, e.g. KNbO$_{3}$ where the potassium-oxygen
 separation in the highest and lowest symmetry structures are about 2.85 \AA~
 and 2.83\AA~, respectively, compared with the sum of their ionic radii which
 is about 2.78 \AA~.

 It is interesting to note that when doping KTaO$_{3}$ with lithium atoms
 (KLT), the system does displace off-center, with a critical concentration of
 lithiums as small as 2.2\% \cite{toulouse}.  This could be the result
 of the lithium ion having a much smaller ionic radii then the potassium
 with respect to the perovskite structure, being about 0.6 \AA~ for Li
 and 1.4\AA~ for K \cite{ionicrad}.  This would allow the lithiums, driven
 by Madelung forces, to displace off-center, and due to the large space open
 to the lithiums in the perovskite structure their amplitudes
 will be much larger then in the LiTaO$_{3}$ system, resulting in a dipole
 field that polarizes the oxygens and distorts them into off-center
 positions. The phase transition in KLT is significantly different then
 in a conventional ferroelectric and there is some discussion of whether
 KLT is a true ferroelectric. DiAntonio {\it et al.}
 \cite{toul2} concluded that the coincidence of the temperature of the maximum
 of the dielectric permittivity with the appearance of other anomalies that
 are characteristics of a structural transformation are a sign that the
 transition is ferroelectric, whereas  Azzini {\it et al.} \cite{azzini}
 state that the size of the domains having a homogeneous spontaneous
 polarization is significantly smaller than the size of the structural
 domains.

 In the Slater picture of the so called ``rattling ion'',
 the B atom lies off center because it is too small to fit into
 the oxygen octahedra surrounding it.  This is in fact the opposite of
 the picture in the LiNb(Ta)O$_{3}$ systems where the separation of the oxygens
 from the Nb (Ta) atoms are smaller (1.9\AA) then the sum of their
 ionic radii (2\AA). Also, a comparison of the Ta-O
 distances in KTaO$_3$ and LiTaO$_3$ shows exactly this same effect. In
 KTaO$_3$, the oxygen octahedra is {\it larger} than the oxygen octahedra in
 LiTaO$_3$, and yet, in KTaO$_3$ the B atom never displaces to the off
 center position, while in LiTaO$_3$, the B cation exhibit a ferroelectric
 distortion.

 \section{Conclusions}

  It is shown that LiNbO$_{3}$ and LiTaO$_{3}$ are very similar in both their
 electronic structure and energetics. The differences in their well depth are
 very small, the amount of
 hybridization in the two materials is similar and the charge densities are
 similar. Also, these two materials hardly change their electronic structure
 during a phase transition.  The only difference found between these two
 systems is  the difference in the conduction bands.
 Zone boundary effects which are not included in this study and this
 difference in the electronic structure of the two systems are
 two possible candidates to explain the difference in the
 transition temperatures of the two systems.

 It is demonstrated  that contrary to previous models which emphasized
 the hopping of the lithium atoms between the three positions as the driving
 mechanism for the phase transformation,  in these systems, no triple
 well potential was found for the lithium motion.
 The deep double wells found are the result of the oxygen displacements
 towards the transition metal atoms, which are the result of the hybridization
 between the two atoms.  The wells indicate an order-disorder character for
 the oxygen.  Local changes in the oxygen octahedra are responsible for
 the lithium displacements from their centrosymmetric sites.  The lithiums
 themselves are passive players in the ferroelectric energetics.

 \underline{Acknowledgements}
 We thank H. Krakauer and D. Singh for the use of the LAPW+LO codes and
 to H. Krakauer, D. Singh, K. Rabe and I. Mazin for helpful discussions.
 This research was supported by the office of Naval Research grant number
 N00014-91-J-1227.

\begin{table}
\caption{Total Energies for different configurations. a) LiNbO$_3$.
 b) LiTaO$_3$. Coordinates are in primitive rhombohedral
 coordinates. The first column refers to the displaced atom and the amount of
 displacement as a fraction of the paraelectric to experimental ferroelectric
 normal mode amplitude.}
 \underbar{  a.) LiNbO$_3$}
 \begin{tabular}{||l|c|c|r||}
 Atomic positions & Li & O & E (Ryd+16193) \\ \hline
 Paraelectric & 0.25,0.25,0.25 & 0.136,0.363,-0.25 & -0.8462 \\
 & & & -0.8464\tablenotemark[1] \\
 Li - 0.5 & 0.2645,0.2645,0.2645 & 0.136,0.363,-0.25 & $-0.8465$ \\
 Li - 1.0 & 0.279,0.279,0.279 & 0.136,0.363,-0.25 & $-0.8467$ \\
 Li - 1.55 & 0.295,0.295,0.295 & 0.136,0.363,-0.25 & $-0.8422$ \\
 O - 0.75 & 0.25,0.25,0.25 & 0.1231,0.3624,-0.2687 & -0.8502 \\
 O - 1.05 & 0.25,0.25,0.25 & 0.1179,0.3621,-0.2761 & -0.8498 \\
 O - 1.3 & 0.25,0.25,0.25 & 0.1135,0.3615,-0.2874 & -0.8473 \\
 O+Li - 0.5 & 0.2645,0.2645,0.2645 & 0.1276,0.3629,-0.2625 & $-0.8539$ \\
 O+Li - 0.75 & 0.272,0.272,0.272 & 0.1231,0.3624,-0.2687 & $-0.8586$ \\
 O+Li - 1.0\tablenotemark[2] & 0.279,0.279,0.279 & 0.1188,0.3622,-0.2749 &
-0.8638 \\
 & & & -0.8644\tablenotemark[1] \\
 O+Li - 1.05 & 0.2804,0.2804,0.2804 & 0.1179,0.3621,-0.2761 & $-0.8643$ \\
 O+Li - 1.3 & 0.288,0.288,0.288 & 0.1135,0.3618,-0.2834 &
$-0.8639$\tablenotemark[3] \\
 O+Li - 1.5 & 0.294,0.294,0.294 & 0.1099,0.3615,-0.2874 &
$-0.8579$\tablenotemark[4] \\
 \end{tabular}

 \underbar{  b.) LiTaO$_3$}
 \begin{tabular}{||l|c|c|r||}
 Atomic positions & Li & O & E (Ryd+ 63395) \\ \hline
 Paraelectric & 0.25,0.25,0.25 & 0.136,0.363,-0.25 & -0.07834 \\
 & & & $-0.07830$\tablenotemark[1] \\
 Li - 0.5 & 0.2645, 0.2645,0.2645 & 0.136,0.363,-0.25 & $-0.0784$ \\
 Li - 1.0 & 0.279,0.279,0.279 & 0.136,0.363,-0.25 & $-0.0778$ \\
 Li - 1.55 & 0.295,0.295,0.295 & 0.136,0.363,-0.25 & $-0.0716$ \\
 O - 0.75 & 0.25,0.25,0.25 & 0.1231,0.3624,-0.2687 & -0.0843 \\
 O - 1.05 & 0.25,0.25,0.25 & 0.1179,0.3621,-0.2761 & -0.0806 \\
 O - 1.3 & 0.25,0.25,0.25 & 0.1135,0.3615,-0.2874 & -0.0719 \\
 O+Li - 0.5 & 0.2645,0.2645,0.2645 & 0.1276,0.3629,-0.2625 & $-0.0898$ \\
 O+Li - 0.75 & 0.272,0.272,0.272 & 0.1231,0.3624,-0.2687 & $-0.0939$ \\
 O+Li - 1.0\tablenotemark[2] & 0.279,0.279,0.279 & 0.1188,0.3622,-0.2749 &
$-0.0957$ \\
  & & & -0.0960\tablenotemark[1] \\
 O+Li - 1.05 & 0.2804,0.2804,0.2804 & 0.1179,0.3621,-0.2761 & $-0.0954$ \\
 O+Li - 1.3 & 0.288,0.288,0.288 & 0.1135,0.3618,-0.2834 &
$-0.0914$\tablenotemark[5] \\
 O+Li - 1.5 & 0.294,0.294,0.294 & 0.1099,0.3615,-0.2874 &
$-0.0809$\tablenotemark[6] \\
 \end{tabular}
\tablenotetext[1]{Calculated using 28 k-points in the irreducible Brillouin
 zone to test convergence.  Other points included 10 k-points.}
\tablenotetext[2]{This is the experimental ferroelectric disortion.}
\tablenotetext[3]{Energy shift due to different muffin tin sizes is
 included as described in the text. Energy shift is 6.64 mRy.}
\tablenotetext[4]{Energy shift due to different muffin tin sizes is 6.6 mRy.}
\tablenotetext[5]{Energy shift due to different muffin tin sizes is 7.14 mRy.}
\tablenotetext[6]{Energy shift due to different muffin tin sizes is 9.68 mRy.}
 \label{ene}
\end{table}

\mediumtext
\begin{table}
\caption{Bond length, in \AA, of transition metals-oxygens and Li-oxygens in
different configurations. Sum of ionic radii of each pair is about 2 \AA.}
 \underbar{a.) LiTaO$_3$}
 \begin{tabular}{||l|c|c|c|c||}
 & Paraelectric & Oxygen Only Distortions & Oxygen Only Distortions &
Ferroelectric \\
 & & Along c axis Only & Along soft-mode coordinate & (Li+O) \\ \hline
 Li-O & 1.99 & 2.00 & 1.96 & 2.04 \\
 Ta-O & 1.97 & 1.86 & 1.91 & 1.91 \\ \hline
 \end{tabular}
\centerline{ }
  \underbar{b.) LiNbO$_{3}$}
 \begin{tabular}{||l|c|c|c|c||}
 & Paraelectric & Oxygen Only Distortions & Oxygen Only Distortions &
Ferroelectric \\
 & & Along c axis Only & Along soft-mode coordinate & (Li+O) \\ \hline
 Li-O & 1.99 & 2.01 & 1.96 & 2.07 \\
 Nb-O & 1.97 & 1.83 & 1.89 & 1.89 \\ \hline
 \end{tabular}
 \label{bonds}
 \end{table}

\narrowtext
\begin{table}
\caption{Parameter table for the polynomial fit of the LAPW and Madelung
 energies. Energies are in Ryd. LAPW LiNbO$_{3}$ energies are shifted by
 -16193 Ryd and LAPW LiTaO$_{3}$ energies by -63395 Ryd.}
  \underbar{a.) LiNbO$_{3}$}
 \begin{tabular}{||l|c|c||}
 Coefficient & LAPW energies  & Madelung energies \\ \hline
   Const &  -0.8467(3) & -31.7268(0) \\
   $Q_{Li}^{2}$ & -0.0005(4) & 0.0194(0) \\
   $Q_{O}^{2}$ & -0.0075(9) & -0.0857(1) \\
   $Q_{Li}Q_{O}$ & -0.0115(8) & 0.0040(0) \\
   $Q_{Li}^{2}Q_{O}^{2}$ & 0.0017(3) & \\
   $Q_{Li}^{4}$ & 0.0003(1) & \\
   $Q_{O}^{4}$ & 0.0038(5) & \\
   & $R^{2}$ = 0.997 & $R^{2}$ = 0.999 \\
 \end{tabular}
\centerline{ }
  \underbar{b.) LiTaO$_{3}$}
 \begin{tabular}{||l|c|c||}
 Coefficient & LAPW energies & Madelung energies \\ \hline
   Const & -0.0800(11) & -31.7714(0) \\
   $Q_{Li}^{2}$ & 0.0009(14) & 0.0200(8)  \\
   $Q_{O}^{2}$ & -0.0117(30) & -0.0812(13) \\
   $Q_{Li}Q_{O}$ & -0.0121(30) & 0.0030(8) \\
   $Q_{Li}^{2}Q_{O}^{2}$ & 0.0010(11) & \\
   $Q_{Li}^{4}$ & 0.0001(3) &  \\
   $Q_{O}^{4}$ & 0.0090(2) &  \\
 & $R^{2}$ = 0.968 & $R^{2}$ = 0.996 \\
 \end{tabular}
 \label{coef}
 \end{table}

 \begin{figure}
 \vspace{1in}
 \caption{The a) paraelectric and b) ferroelectric structures of LiTaO$_3$ and
 LiNbO$_{3}$. The hexagonal unit cell is outlined.}
 \label{structure}
 \end{figure}

 \begin{figure}
 \vspace{1in}
 \caption{a.) Potential energy surfaces of LiTaO$_3$.  The
upper curves represent  displacements of the Li atoms along the soft-mode
 coordinate, the middle shallow double wells represent displacements
of the oxygens alone and the bottom curves represent the displacements of
oxygens and Li atoms along the same coordinate. The curves represent
a 4th order polynomial fit to the data. They were not constrained to
go through the zero of energy. The abscissa represents displacement of
the oxygen atoms from the paraelectric configuration, in \AA. b.) The same
 for LiNbO$_{3}$.}
 \label{energy}
 \end{figure}

 \begin{figure}
 \vspace{1in}
 \caption{The energy as a function of normal coordinate,
  fitted to a quadratic.  The lines are the eigenstates for
  the 1-D independent harmonic oscillators. The difference between the ground
 state and the lowest excited state gives a frequency of 270 $cm^{-1}$ for
 LiTaO$_{3}$ and 250 $cm^{-1}$ for LiNbO$_{3}$. Both are in very good
 agreement with experimental results.  a) Energy versus normal coordinate for
 LiTaO$_3$ and b) The same for LiNbO$_3$.}
 \label{levels}
 \end{figure}

 \begin{figure}
 \vspace{1in}
 \caption{Total energy surfaces of LiNbO$_{3}$ with only oxygens
 displaced along soft mode coordinate (lower curve) and with only the oxygens
 displaced along the c axis only (upper curve). The curves are a 4th order fit
 to the data. The abscissa represents displacements, in \AA, of the oxygens
 from their paraelectric positions.}
 \label{zonly}
 \end{figure}

 \begin{figure}
 \vspace{1in}
 \caption{LiNbO$_{3}$: Charge density resulting from subtracting
 the charge density of a configuration in which oxygens only are displaced
 from the charge density of the full ferroelectric distortion (both lithiums
 and oxygens are displaced).  The scale is from -0.1 to 0.1 electrons per
 $bohr^{3}$ and the contour interval is 0.002 electrons/bohr$^{3}$. No
 evidence for dynamical covalency effects that would lead to coupling of
 oxygen and lithium motions are seen}
 \label{polarization}
 \end{figure}

 \begin{figure}
 \vspace{1in}
 \caption{a) Electronic density of LiTaO$_{3}$ (solid line) and LiNbO$_{3}$
 (dashed line), both in the experimental ferroelectric configuration.
 b) Density of Ta 5d states (solid line) and Nb 4d states (dashed line) in
 the valence band, both the experimental ferroelectric configuration.
 c) Oxygen p states of LiTaO3 (solid
 line) and of LiNbO$_{3}$ (dashed line) in the valence band, in the
 ferroelectric phase.}
 \label{dos2}
 \end{figure}

 \begin{figure}
 \vspace{1in}
 \caption{a) Electronic density of states for LiNbO3 in the paraelectric
 (solid line) and ferroelectric (dashed line) configurations.
 b) The Nb 4d state in the two phases. c) The LiTaO$_{3}$ Ta 5d states in
 the two phases. d)  The LiNbO$_{3}$'s p states on the oxygens in the
 paraelectric and the ferroelectric phase.  e) The s orbital on the Li atom
 of LiNbO$_{3}$ in the two phases.  The valence band top is lined up with
 the zero of energy.}
 \label{dos1}
 \end{figure}

 \begin{figure}
 \vspace{1in}
 \caption{  The band structure of LiNbO$_{3}$ in the ferroelectric phase.
 The energy scale is in eV and the fermi level is shown.  The band gap is
 3.1 eV.   The Li s states do not interact with the rest of the bands and
 are about 16 eV below the Oxygen 2s states. Inset: The brillouin zone.
 Some high symmetry points are illustrated. The energy between the Z and the
 A point was calculated along a straight line between the two points.
 From [60].}
 \label{bands}
 \end{figure}

 \begin{figure}
 \vspace{1in}
 \caption{a)  The band structure of LiTaO$_{3}$ in the ferroelectric phase
 (solid line) and the paraelectric (dashed line).  Only the valence and the
 conduction bands are shown.  The band gap in the ferroelectric phase is
 about 4.0 eV, and is decreased in the paraelectric phase by about 15 \%.
 b) The same for LiNbO$_{3}$.  The gap decreases by about 15 \% between
 the ferroelectric and paraelectric phases. No major differences are
 observed in both systems between the two phases.}
 \label{bands1}
 \end{figure}

  \begin{figure}
 \vspace{1in}
 \caption{The band structure of LiNbO$_{3}$ (solid line) and LiTaO$_{3}$
 (dashed line), both in the ferroelectric structure.  The LiTaO$_{3}$ band
 gap is larger than the LiNbO$_{3}$ band gap by about 1 eV.
 The valence bands of the two materials are almost identical.}
 \label{bancom}
 \end{figure}

 \begin{figure}
 \vspace{1in}
 \caption{ From LMTO results of KTaO$_{3}$ and KNbO$_{3}$, by Postnikov {\it et
al.} [29].  a) Local densities of states at the Nb site with the Nb
 undisplaced (solid line) and displaced (dashed line) from it's rhombohedral
 position.  The displacement of the Nb atom is exaggerated in order to enhance
 the trends shown.  b) at the Ta site, under the same conditions. The units are
in
 Ryd for the energies and $Ryd^{-1}$ for the density of states.}
 \label{pos}
 \end{figure}

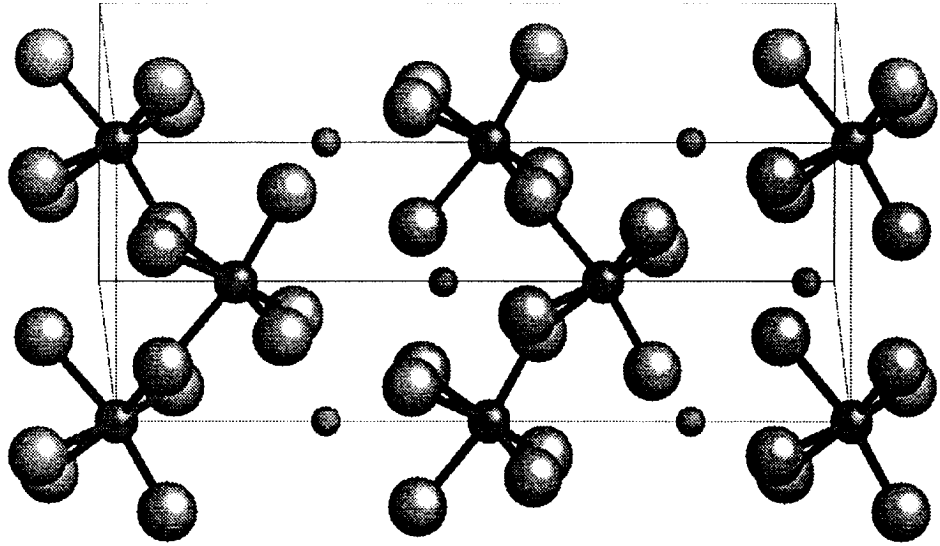
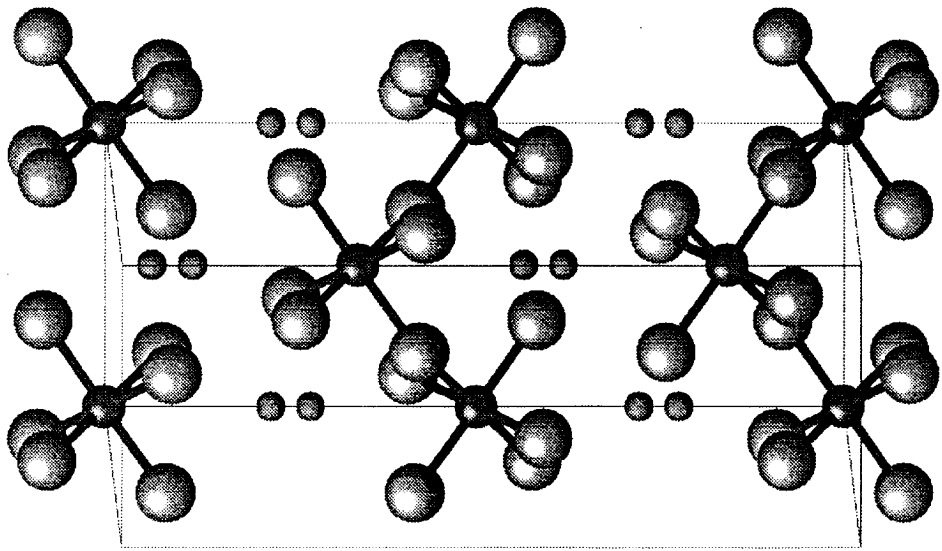
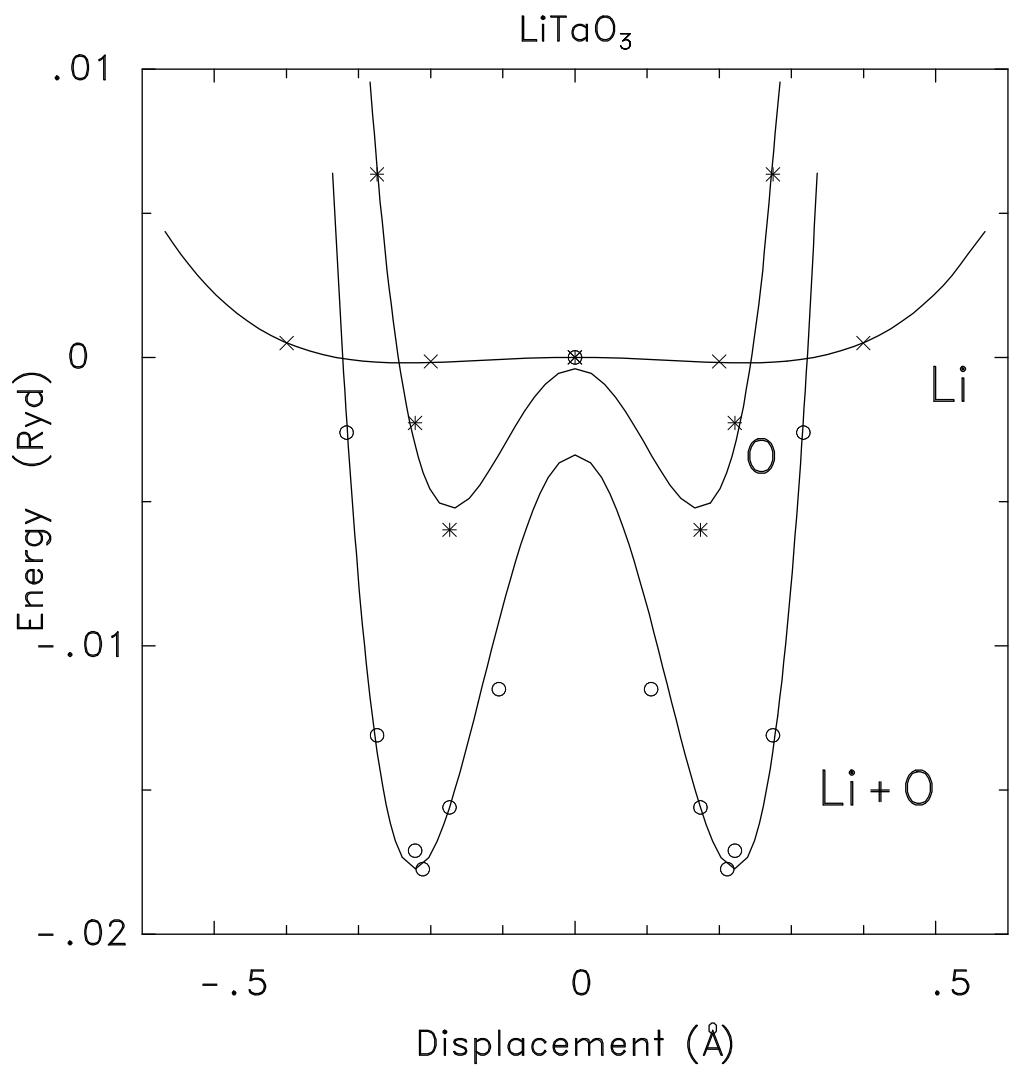
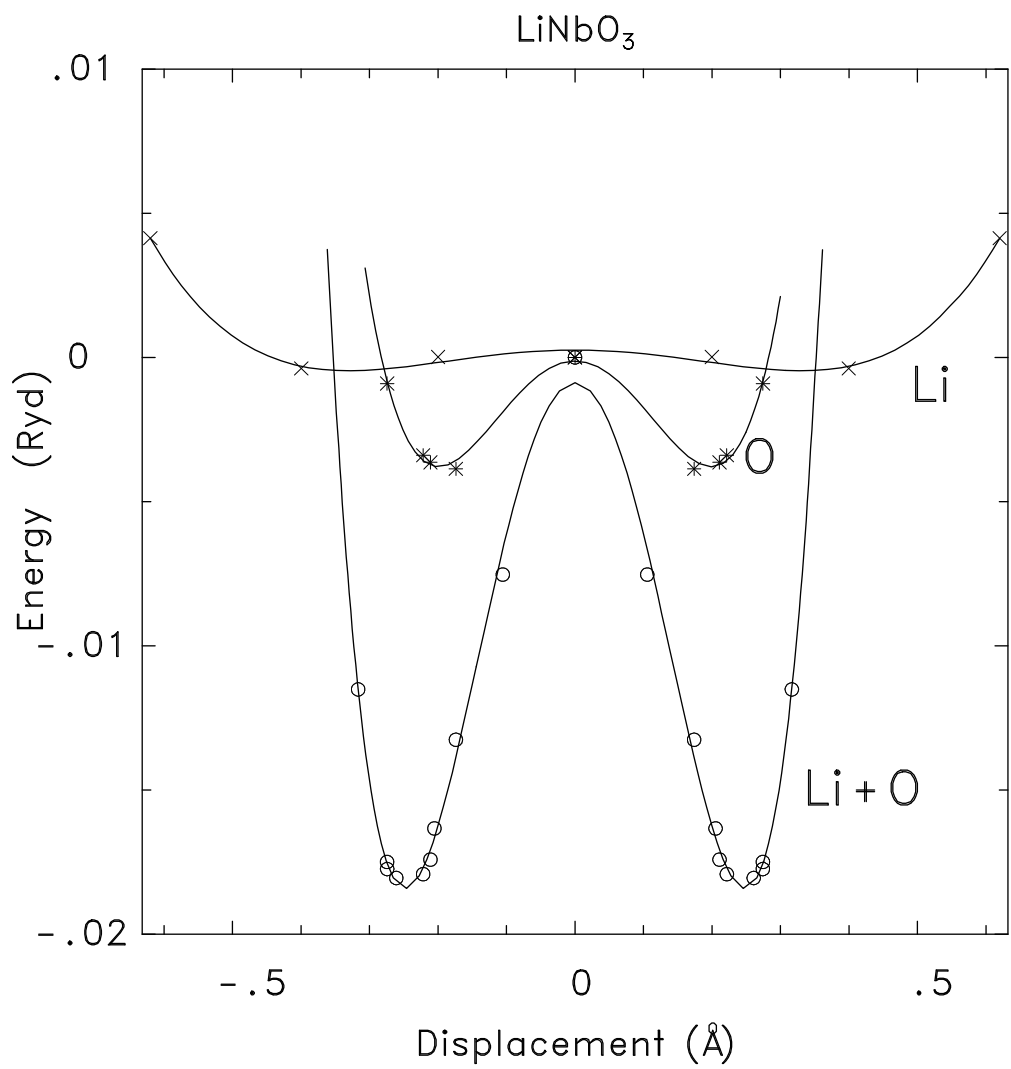
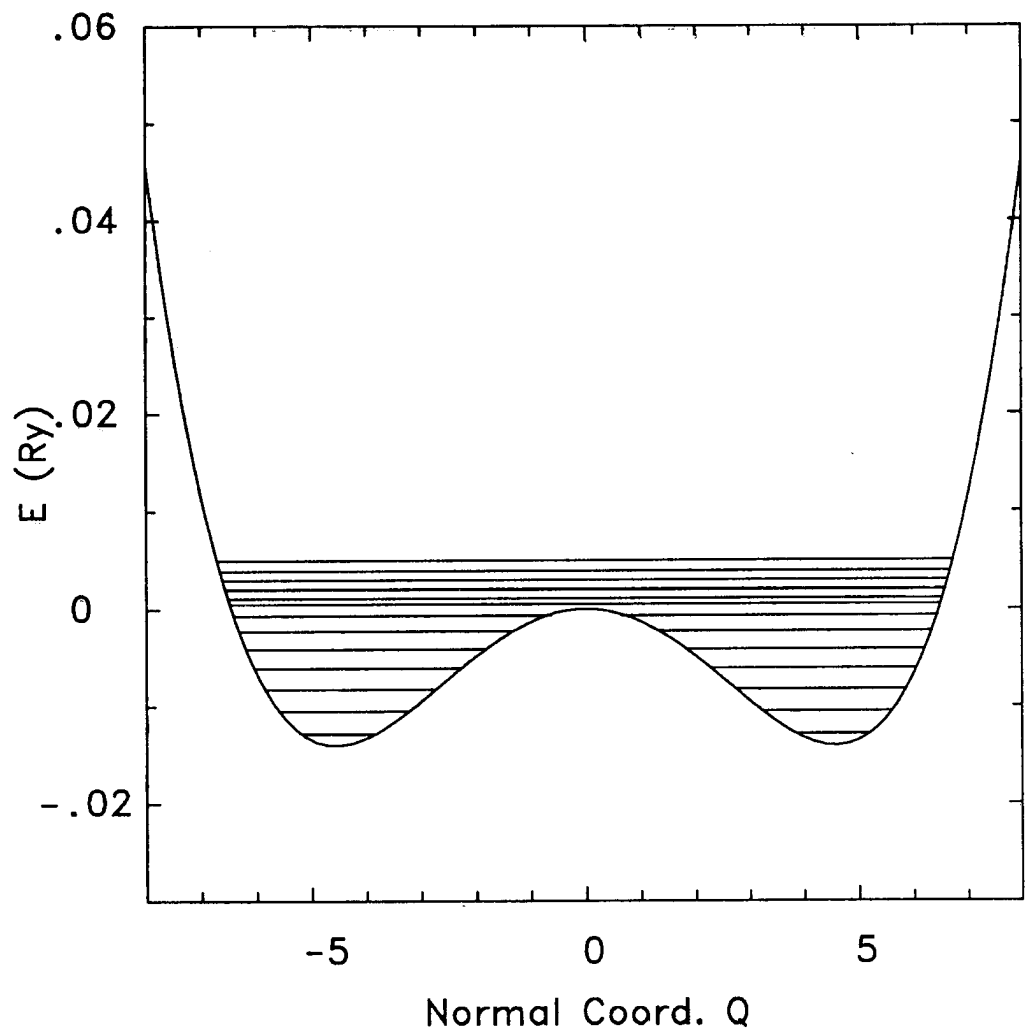
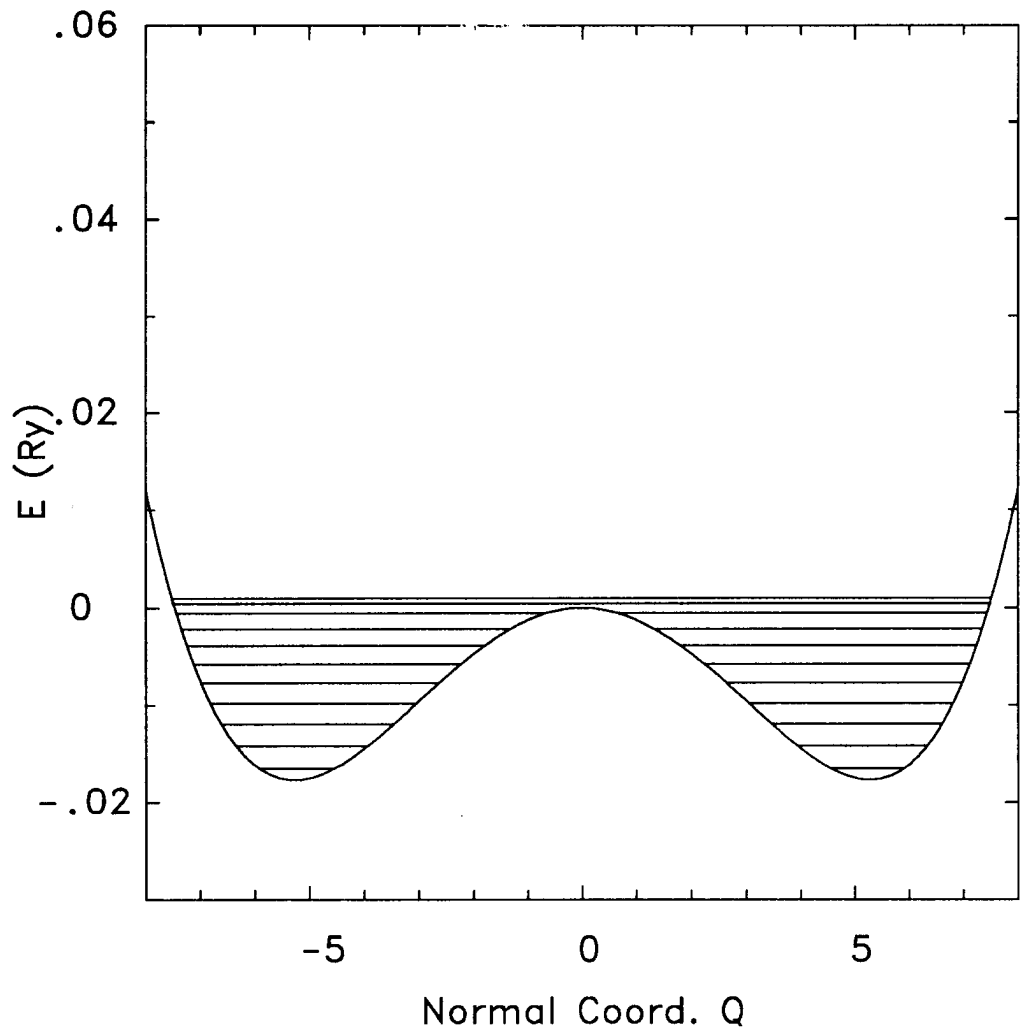
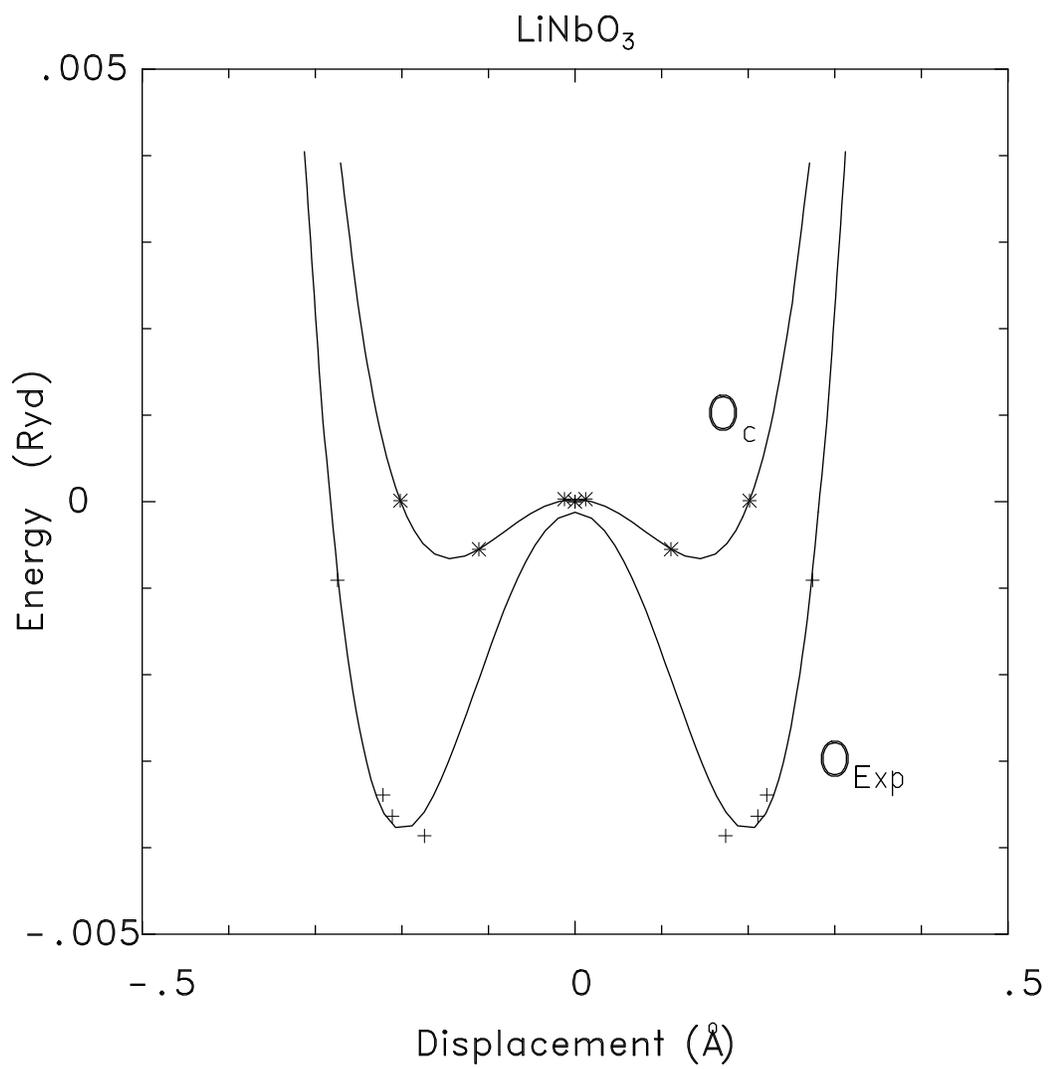
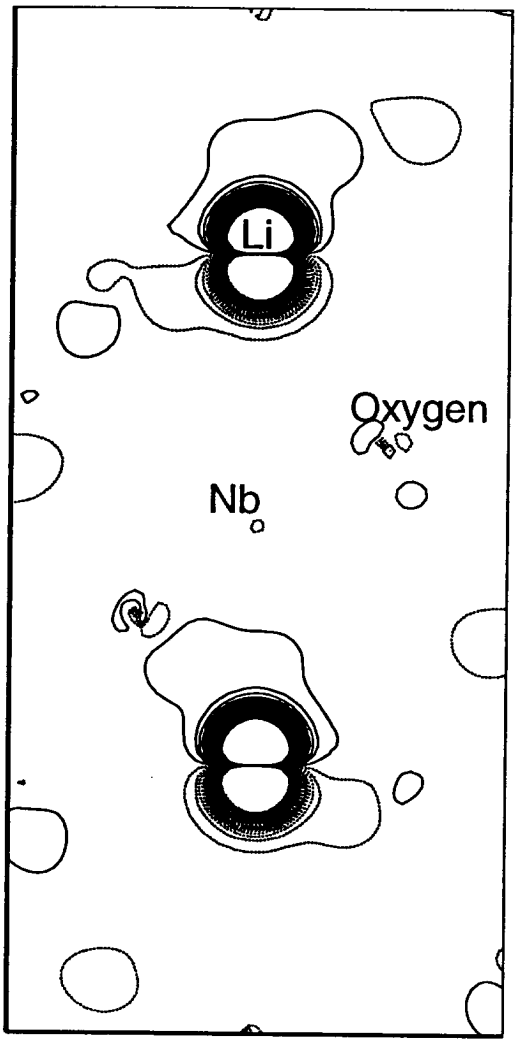
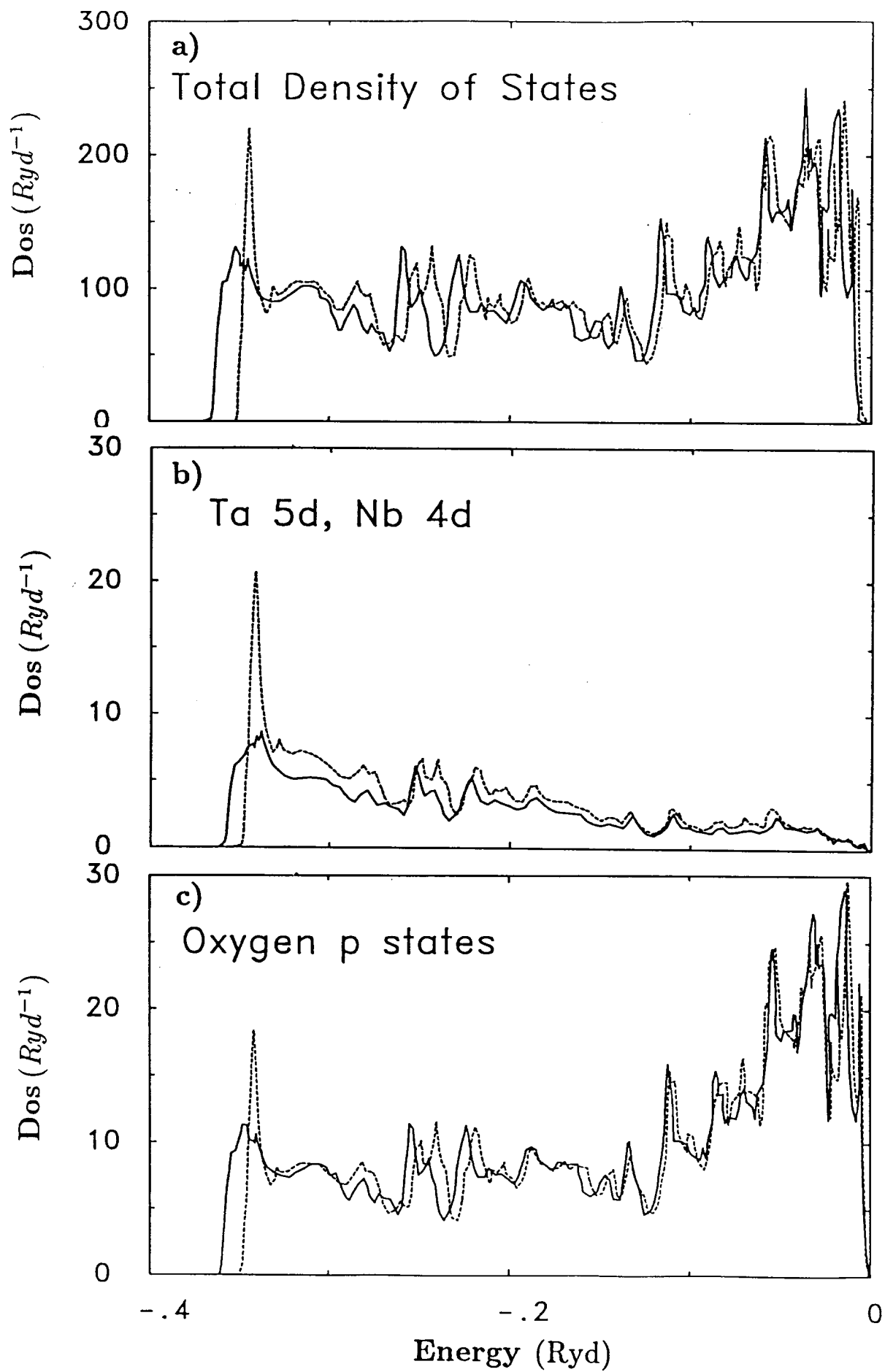
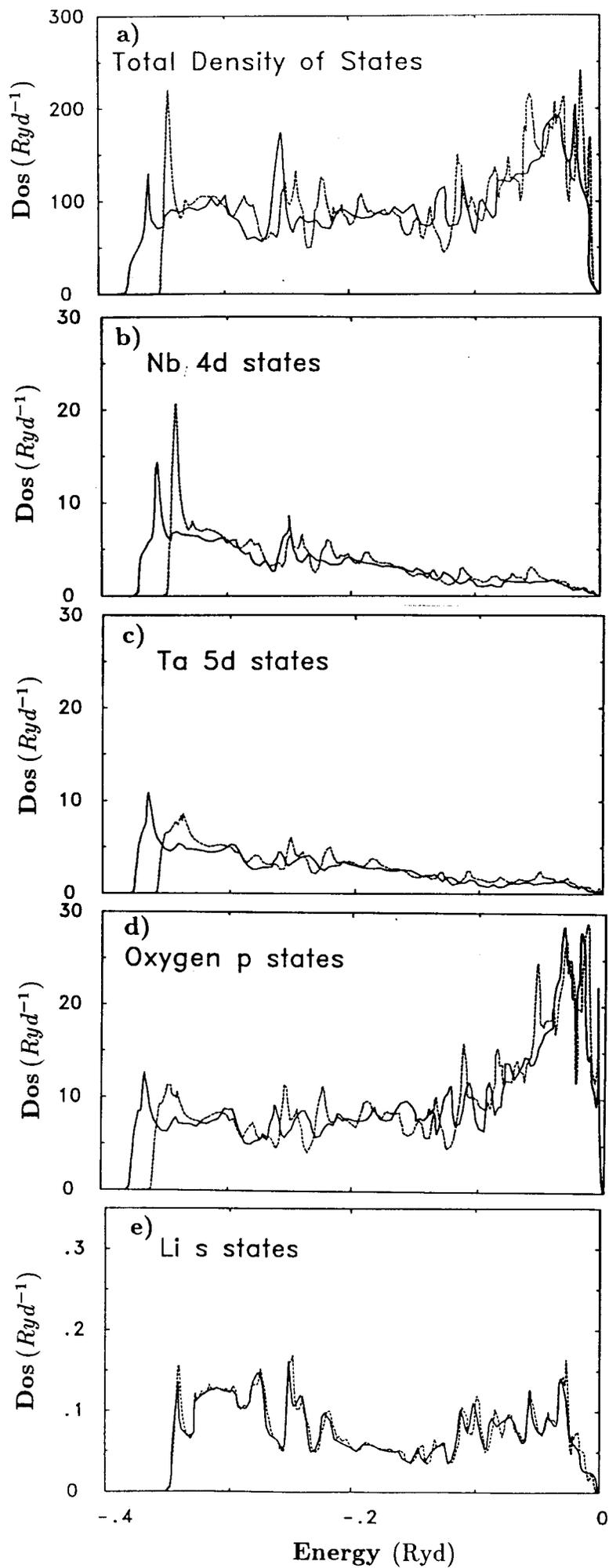
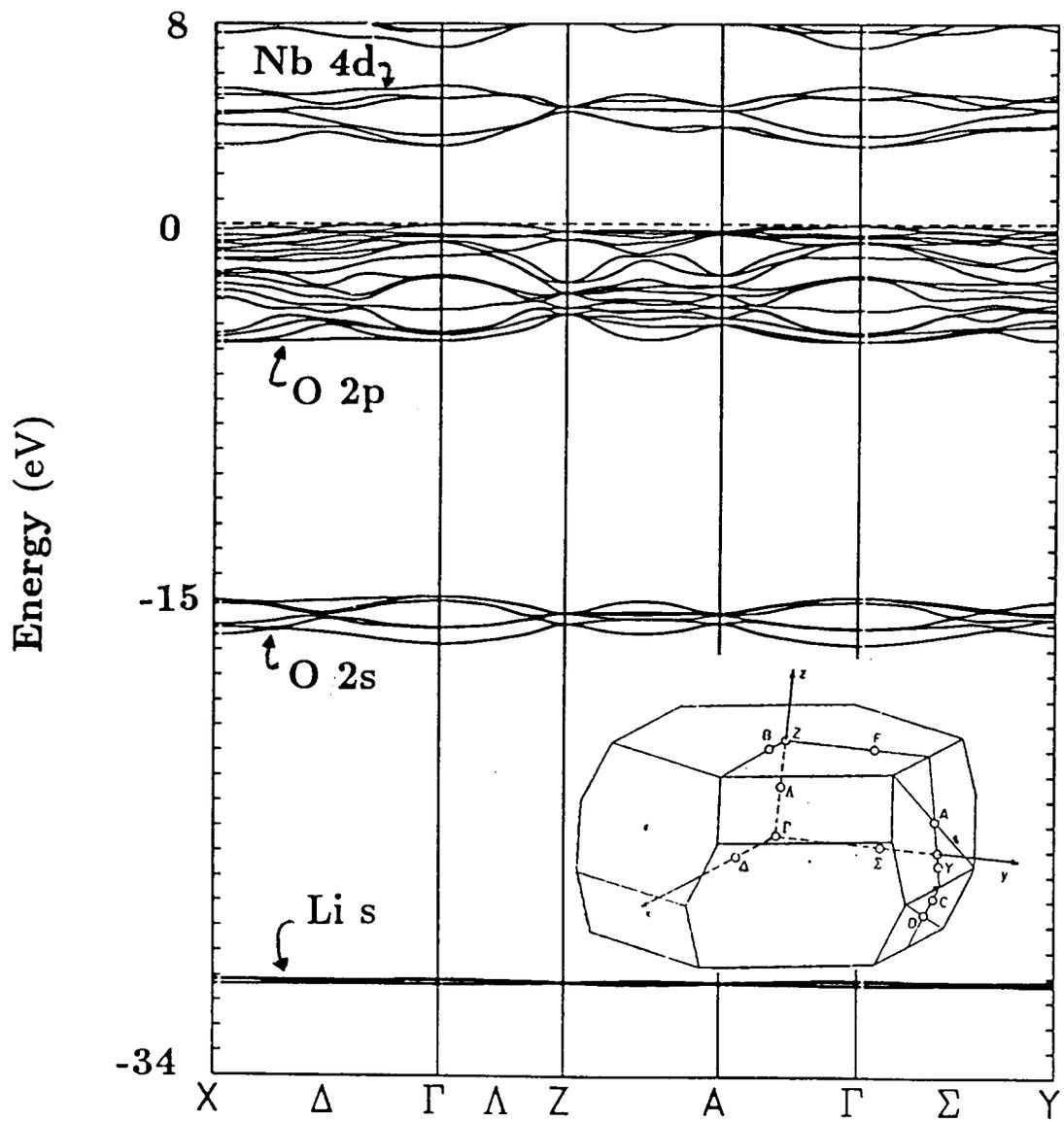
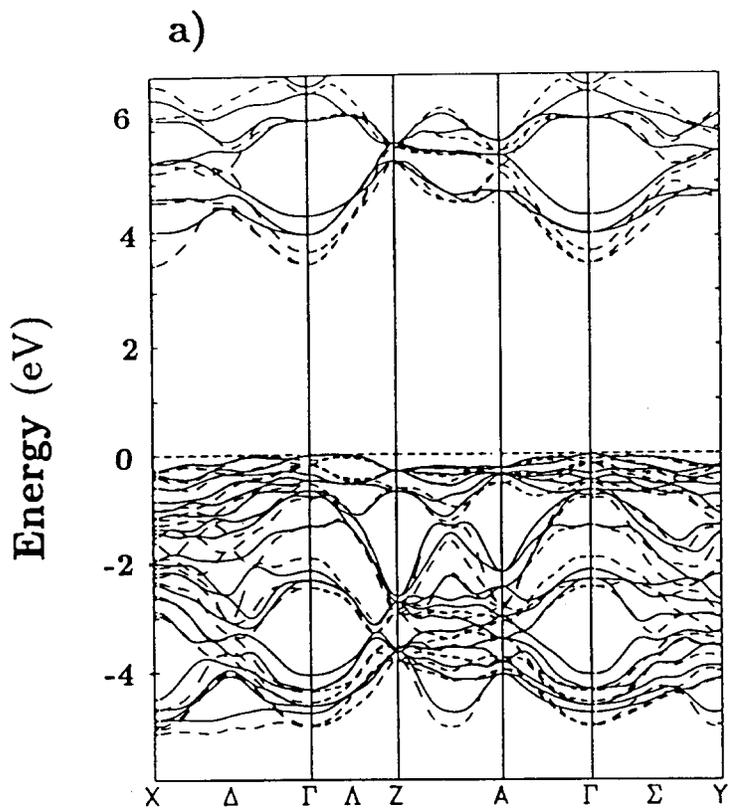
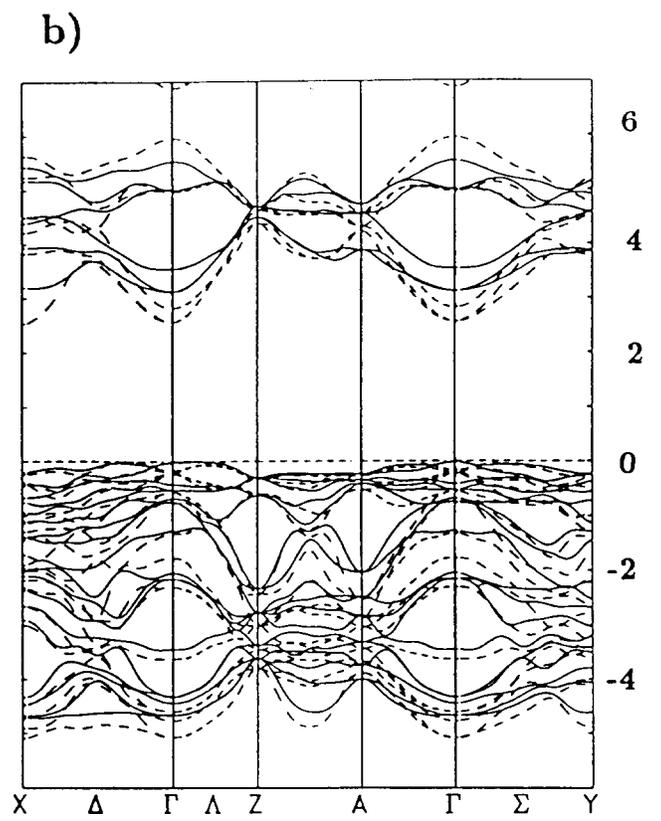
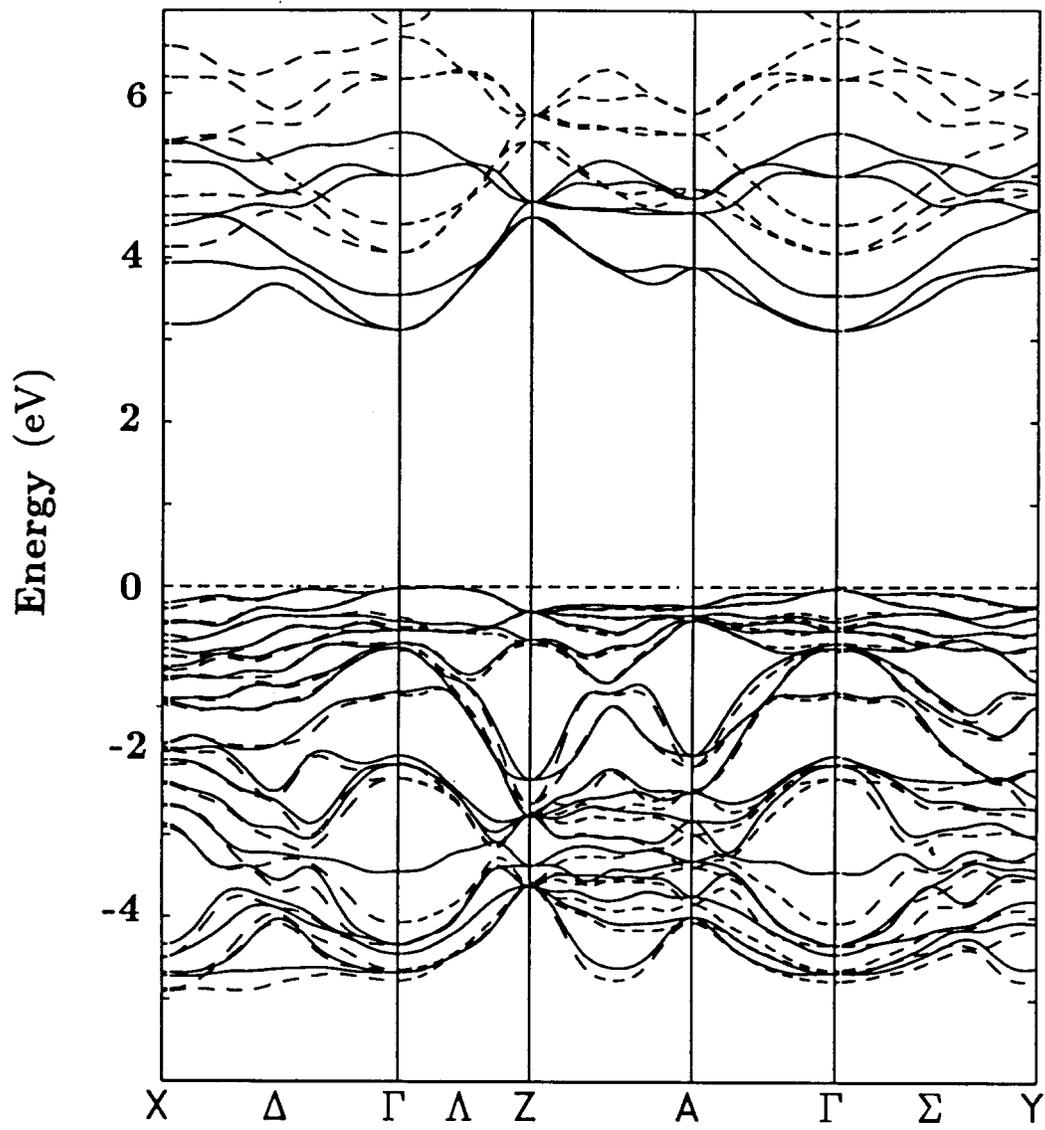
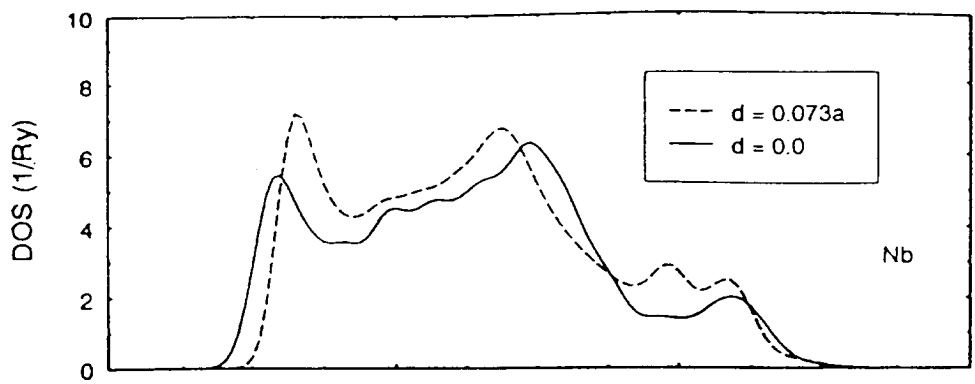
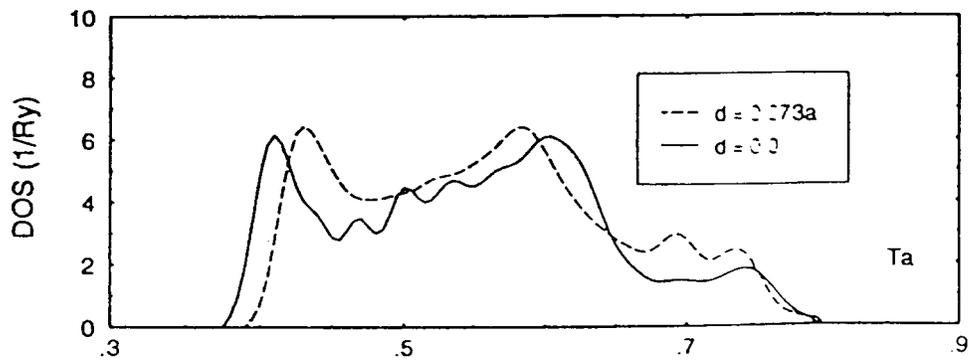
\end{document}